\documentclass[aps,pre,reprint,longbibliography,superscriptaddress]{revtex4-2}
\usepackage{amsmath}
\usepackage{graphicx}
\usepackage{hyperref}

\newcommand{\bra}[1]{\left\langle #1\right|}
\newcommand{\ket}[1]{\left| #1\right\rangle}
\newcommand{\nangle}[1]{\left\langle #1 \right\rangle}

\newcommand{\e}{{\rm e}}
\newcommand{\imai}{{\rm i}}

\usepackage[normalem]{ulem} % for sout{} command
\usepackage{color}
\begin{document}
\title{Violating the Thermodynamic Uncertainty Relation in the Three-Level Maser}
\author{Alex Arash Sand Kalaee, Andreas Wacker}

\author{Patrick P. Potts}
\affiliation{Mathematical Physics and NanoLund, Lund University, Box 118, 221 00 Lund, Sweden}
\affiliation{%
	Department of Physics, University of Basel, Klingelbergstrasse 82, 4056 Basel,
	Switzerland
}%

\date{\today}

\begin{abstract}
Nanoscale heat engines are subject to large fluctuations which affect their precision.
The Thermodynamic Uncertainty Relation (TUR) provides a trade-off between output power, fluctuations and entropic cost.
This trade-off may be overcome by systems exhibiting quantum coherence.
This letter provides a study of the TUR in a prototypical quantum heat engine, the Scovil \& Schulz-DuBois maser.
Comparison with a classical reference system allows us to determine the effect of quantum coherence on the performance of the heat engine.
We identify analytically regions where coherence suppresses fluctuations, implying a quantum advantage, as well as regions where fluctuations are enhanced by coherence.
This quantum effect cannot be anticipated from the off-diagonal elements of the density matrix.
Because the fluctuations are not encoded in the steady state alone, TUR violations are a consequence of coherence that goes beyond steady-state coherence.
While the system violates the conventional TUR, it adheres a recent formulation of a quantum TUR.
We further show that parameters where the engine operates close to the conventional limit are prevalent and TUR violations in the quantum model not uncommon.
\end{abstract}
\maketitle

\textit{Introduction.---}
Nanoscale heat engines \cite{BenentiPhysRep2017} have become a topic of wide interest in recent years. In such devices quantum effects become relevant and radically alter the dynamical and thermodynamic properties \cite{SothmannNano2014,MillenNJP2016,VinjanampathyContPhys2016,StreltsovRevModPhys2017,JosefssonNatNano2018,BinderBook2018,TalknerRevModPhys2020}.
Nanoscale systems are subject to strong fluctuations in the output power $P$ which become important when quantifying useful properties such as efficiency and precision \cite{SeifertRepProgPhys2012,VerleyNatComm2014}.
At the same time, operation of any engine is associated with the entropy production rate $\sigma$, which quantifies the thermodynamic cost. It is desirable for an engine 
to have both a low entropy production rate and a high precision.
Their trade-off can be quantified by the dimensionless thermodynamic uncertainty
\begin{equation}
\mathcal{Q}=\frac{\sigma}{k_B}\frac{\mathrm{var}(P)}{\nangle{P}^2},
\label{EqTUR}
\end{equation}
where $\langle P\rangle$ and $\mathrm{var}(P)$ denote the mean and variance of the power in the long time limit and $k_B$ is the Boltzmann constant. The Thermodynamic Uncertainty Relation (TUR) provides a lower bound $\mathcal{Q}\ge 2$ \cite{BaratoPRL2015,GingrichPRL2016,PietzonkaJStatMech2016,PietzonkaPRE2017,
PietzonkaPRL2018,HorowitzNatPhys2019,FalascoNJP2020}, which highlights the fundamental relevance of the thermodynamic uncertainty $\mathcal{Q}$.

The TUR has been applied to biomolecular processes \cite{BaratoPRL2015,PietzonkaJStatMech2016,JackPRE2020,SongJPhysChemLett2020}, heat transport \cite{SaryalPRE2019}, and Brownian clocks \cite{BaratoPRX2016}.
Furthermore, experimental realizations are an area of active development \cite{ManikandanArxiv2021,PaneruPRE2020,HwangJPhysChemLett2018,PalPRR2020,FriedmanPRB2020}.
Generalizations have been formulated to cover discrete or time-dependent driving \cite{BaratoPRX2016,ProesmansEPL2017,macieszczak:2018,KoyukPRL2020,cangemi:2021}, underdamped Langevin dynamics \cite{FischerPRE2018,LeePRE2019,VuPRE2019}, and the presence of measurement and feedback \cite{PottsPRE2019}.
Beyond the classical regime, quantum effects influence work, fluctuations, and entropy production \cite{KiesslichPRB2006,ScandiPRR2020,LatunePRA2020} altering the TUR bounds on quantum stochastic systems \cite{BrandnerPRL2018,CarolloPRL2019,GuarnierePRR2019,PalPRR2020,HasegawaPRL2020,HasegawaPRL2021}, e.g., due to coherences \cite{PtaszynskiPRB2018,AgarwallaPRB2018,cangemi:2020,RignonBretPRE2021} or particle exchange correlations \cite{BrandnerPRL2018,SaryalPRE2019,LiuPRE2019}.
Understanding the detailed reasons for such TUR violations and the possible establishment of more general bounds are important questions in ongoing research.

The Scovil \& Schulz-DuBois (SSDB) three-level maser is the prototype for quantum heat engines relying on quantum coherence to do work \cite{ScovilPRL1959}.
In this work we study the thermodynamic uncertainty in the SSDB maser in detail, finding TUR violations induced by coherence in analogy to Ref.~\cite{PtaszynskiPRB2018}.
A comparison to a classical model, which obeys the TUR allows us to
identify regions of operation where quantum dynamics results in an improved operation as quantified by a lower value of $\mathcal{Q}$. 
Interestingly, such a quantum advantage cannot be anticipated from the off-diagonal elements of the density matrix because the fluctuations are not encoded in the steady state alone.
TUR violations should thus be seen as a consequence of coherent dynamics going beyond steady-state coherence.

\begin{figure}
\centering
\includegraphics[width=\columnwidth]{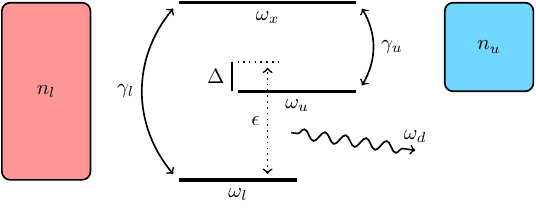}
\caption{Illustration of the three level maser with energies $\omega_\alpha$ for $\alpha\in\{x,u,l\}$.
Bath $\alpha$ with population $n_\alpha$ induces transition with rate $\gamma_\alpha$.
The external ac field connects the levels $u$ and $l$ with strength $\epsilon$ and frequency $\omega_d = \Delta + \omega_u-\omega_l$.
In one cycle of operation, the system is excited from bath $l$ to an excited state $x$, from which it relaxes to the upper laser level $u$ by emitting heat to bath $u$.
The cycle is closed by a photon that is emitted into the driving field, producing work.
}
\label{FigSchema}
\end{figure}

\begin{figure*}
	\centering
	\includegraphics[width=\textwidth]{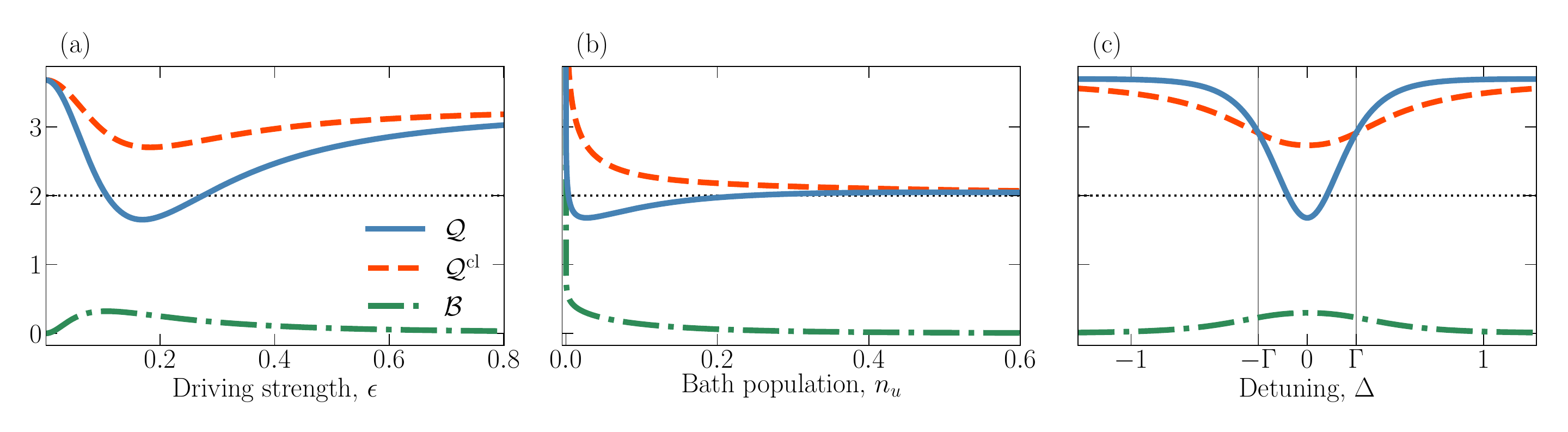}
	\caption{Probe of thermodynamic uncertainty for the SSDB maser $\mathcal{Q}$ (solid blue) and the classical reference system $\mathcal{Q}^\mathrm{cl}$ (dashed orange) along (a)  driving strength $\epsilon$, (b) bath population $n_u$ and (c) detuning $\Delta$. The fixed parameters are $\gamma_u=2$, $\gamma_l = 0.1$ and $n_l=5$, while the parameters $n_u = 0.027$, $\epsilon=0.15$ and $\Delta=0$ are varied according to the respective panels.
		The classical TUR limit is indicated by the dotted black line and the quantum TUR bound $\mathcal{B}$ with dash-dotted green. 
	}
	\label{FigProbe}
\end{figure*}

\textit{Quantum advantage in the maser.---}
Figure \ref{FigSchema} shows the maser \cite{ScovilPRL1959}, where the three levels are the lower $l$ and upper $u$ lasing levels, and the excited state $x$.
If $n_l>n_u$, where $n_\alpha$ is the relevant occupation of the Bosonic bath coupled to level $\alpha$ with energy $\omega_\alpha$, we obtain inversion between the laser levels, resulting in maser operation.
The steady-state work output of this type of engine is well known from earlier analysis in interaction with classical \cite{LambPR1964,GevaPRE1994,GevaJChemPhys1996,BoukobzaPRA2006,KalaeePRA2021} or quantized \cite{NiedenzuQuantum2019} light.

We probe the thermodynamic uncertainty of the SSDB maser $\mathcal{Q}$ and compare it with the uncertainty of an equivalent classical system $\mathcal{Q}^\mathrm{cl}$, where the coherent transition between $u$ and $l$ is replaced by a classical rate, see Fig.~\ref{FigProbe}.
While the classical system always adheres the TUR, we find regions where $\mathcal{Q}$ can go as low as 1.68, a significant violation enabled by the quantum-coherent dynamics of the maser.
These violations occur at intermediate driving strength $\epsilon$, cf. Fig.~\ref{FigProbe}~(a), while the classical model captures the behavior of the quantum model for both weak and strong driving.
For small $\epsilon$ a perturbative treatment of the drive is justified while for large $\epsilon$, the statistics is determined by the rates which mediate heat transfer with the baths, $\gamma_l$ and $\gamma_u$.
Figure \ref{FigProbe}~(b) shows that the most significant TUR violation occurs at low population $n_u$, where the associated temperature and decoherence are low.
Interestingly, coherence may also be disadvantageous, i.e., $\mathcal{Q} > \mathcal{Q}^\mathrm{cl}$.
This disadvantage happens at finite detuning $|\Delta| > \Gamma$, where $\Gamma=(\gamma_un_u+\gamma_ln_l)/2$ denotes the broadening of the transition (which is equal to the decoherence rate), see Fig.~\ref{FigProbe}~(c).

Recently, it was suggested that quantum systems which evolve according to Lindblad-type master equations adhere to a quantum TUR \cite{HasegawaPRL2020}
$\mathcal{Q}\geq\mathcal{B}$, (see the supplemental material below for a detailed evaluation of $\mathcal{B}$ for the SSDB).
For the parameters which provided a significant violation of the classical TUR we find that the SSDB maser remains well within adherence to the quantum TUR, see Fig.~\ref{FigProbe}~(a-c).

\textit{The system master equation.---}
For a quantitative description,
the system is assumed to have a Markovian time evolution governed by the Lindblad master equation \cite{LindbladCommMathPhys1976} (we use $\hbar=k_B=1$ in the following)
\begin{equation}
\dot\rho = -\imai[H(t),\rho]+\sum_\alpha\left\{\gamma_\alpha(n_\alpha+1)\mathcal{D}_{\sigma_{\alpha x}}[\rho]+\gamma_\alpha n_\alpha\mathcal{D}_{\sigma_{x\alpha}}[\rho]\right\}\,,
\label{EqSystemLindblad}
\end{equation}
with the dissipator $\mathcal{D}_\sigma[\rho] = \sigma\rho\sigma^\dagger-\frac12\{\sigma^\dagger\sigma\rho+\rho\sigma^\dagger\sigma\}$.
Here we introduced the bath populations $n_\alpha$ and the respective transition rates $\gamma_\alpha$ for $\alpha\in\{u,l\}$.
The Hamiltonian $H(t) = H_0 + V(t)$ consists of a bare term
$H_0 = \omega_l\sigma_{ll}+\omega_u\sigma_{uu}+\omega_x\sigma_{xx}$
and an external classical field
$V(t) = \epsilon(\e^{\imai\omega_dt}\sigma_{lu}+e^{-\imai\omega_dt}\sigma_{ul})$
with frequency $\omega_d$ and strength $\epsilon$.
The transition operators are defined as $\sigma_{ij} = \ket{i}\bra{j}$.

In addition to the quantum SSDB maser we also consider a classical reference system (with a superscript cl described by the master equation
\begin{equation}
\begin{split}
\dot\rho^\mathrm{cl} =& \gamma_c\left(\mathcal{D}_{\sigma_{ul}}[\rho^\mathrm{cl}]+\mathcal{D}_{\sigma_{lu}}[\rho^\mathrm{cl}]\right)\\&+\sum_{\alpha}\left\{\gamma_\alpha(n_\alpha+1)\mathcal{D}_{\sigma_{\alpha x}}[\rho^\mathrm{cl}]+\gamma_\alpha n_\alpha\mathcal{D}_{\sigma_{x\alpha}}[\rho^\mathrm{cl}]\right\}\,,
\label{EqSystemClassical}
\end{split}
\end{equation}
where we have introduced the classical transition rate
\begin{equation}
\gamma_c=\frac{2\epsilon^2\Gamma}{\Delta^2+\Gamma^2}\,,
\label{EqSystemCoupling}
\end{equation}
as obtained by Fermi's golden rule assuming Lorentzian broadening, which provides the same work output as the quantum model.

\textit{The quantities of interest.---}
In order to calculate $\mathcal{Q}$ we need the mean and variance of the power as well as the entropy production rate.
Let $N$ denote the number of cycles done in nominal direction (i.e., $l\rightarrow x\rightarrow u\rightarrow l$) minus the number of cycles in opposite direction.
In each cycle, a photon is exchanged with each heat bath. As shown in the supplemental material below, we further find $\langle P\rangle=\omega_d\langle \dot{N}\rangle$ and
$\mathrm{var}(P)=\omega_d^2\mathrm{var}(\dot{N})$, consistent with the picture that a single photon is emitted into the drive field per cycle. We note that this simple relation between heat and work was observed before \cite{VerteletskyPRA2020}, and that it may break down for higher cumulants \cite{KerremansArxiv2021}.

The entropy production rate reads \cite{SpohnJMathPhys1978}
\begin{equation}
\sigma = -\frac{\dot{Q}_u}{T_u} - \frac{\dot{Q}_l}{T_l}\,,
\end{equation}
where $\dot{Q}_\alpha$ is the net rate of heat transferred from bath $\alpha$ to the engine. $T_\alpha$ is the bath temperature defined from the Bose-Einstein distribution
\begin{equation}
\label{eq:temps}
T_\alpha = \frac{\Omega_\alpha}{\ln\left(1+\frac{1}{n_\alpha}\right)}\,,
\end{equation}
where $\Omega_\alpha=|\dot{Q}_\alpha/\langle\dot{N}\rangle|$ denotes the heat per particle transfer. We note that by using Eq.~\eqref{eq:temps}, our results hold for both the definition of heat and work via the full or bare Hamiltonian, see Ref.~\cite{KalaeePRA2021} for more information on this debate.
This yields a positive definite entropy production rate 
\begin{equation}
\sigma =\ln\left[\frac{n_l(n_u+1)}{n_u(n_l+1)}\right]\langle\dot{N}\rangle > 0,
\label{EqTUREntropy}
\end{equation}
as $\langle\dot{N}\rangle$ has the same sign as the thermodynamic driving $n_l-n_u$, see Eq.~\eqref{EqTURmean}.
Inserting the power and entropy into the definition of $\mathcal{Q}$ we obtain
\begin{equation}
\mathcal{Q} =
\ln\left[\frac{n_l(n_u+1)}{n_u(n_l+1)}\right]F\, ,
\label{EqQturScovil}
\end{equation}
where we introduced the Fano factor $F=\frac{\mathrm{var}(\dot{N})}{\langle\dot{N}\rangle}$.

To determine the mean and variance of $\dot{N}$ we employ Full Counting Statistics (FCS) \cite{BrudererNJP2014,EspositoRevModPhys2009,SchallerBook2014}
where we modify the master equations by introducing counting fields to keep track of the number of energy quanta exchanged with the baths.
This procedure yields the same mean rate for the quantum and the classical model as detailed in the supplemental material below
\begin{equation}
\langle\dot{N}\rangle =
\frac{\gamma_c\gamma_u\gamma_l(n_l-n_u)}{\gamma_u\gamma_l(3n_ln_u+n_u+n_l)+2\gamma_c(3\Gamma+\gamma_u+\gamma_l)}\, .
\label{EqTURmean}
\end{equation}

The Fano factor can be written as
\begin{equation}
F = \frac{n_l(n_u+1)+n_u(n_l+1)}{n_l-n_u} - 2\langle\dot{N}\rangle C\,,
\label{EqFano}
\end{equation}
where $C$ takes the values
\begin{equation}
\begin{split}
C^\mathrm{cl} = &
\frac{2\gamma_c+4\Gamma+\gamma_l+\gamma_u}{D}
\label{EqC}\\
C =&C^\mathrm{cl}+\frac{\Gamma^2-\Delta^2}{\Delta^2+\Gamma^2}\frac{\gamma_u\gamma_l}{\Gamma}
\frac{ \left(3 n_{l} n_{u} + n_{l} + n_{u}\right)}{D}
\end{split}\end{equation}
with $D= \gamma_u\gamma_l(3n_ln_u+n_u+n_l)+2\gamma_c(3\Gamma+\gamma_u+\gamma_l)$
for the classical and the quantum model, respectively.
In contrast to the average rate, the variance thus differs between the quantum and the classical model. Here $C$ is larger (smaller) than
$C^\textrm{cl}$, if the detuning $|\Delta|$ is smaller (larger) than the broadening  $\Gamma$, respectively. From Eqs.~(\ref{EqQturScovil}) and (\ref{EqFano}), we find
\begin{equation}
\mathcal{Q}-\mathcal{Q}^\mathrm{cl}=2\langle\dot{N}\rangle
\ln\left[\frac{n_l(n_u+1)}{n_u(n_l+1)}\right]\left(C^\mathrm{cl}-C\right)\, .
\label{EqQadvantage}
\end{equation}
We find the quantum model results in reduced noise $F$ and thermodynamic uncertainty $\mathcal{Q}$ for small detuning $|\Delta|<\Gamma$. This establishes the condition for a quantum advantage in the SSDB engine, implying that this engine outperforms its classical limit. Furthermore, when $\mathcal{Q}<2$, the SSDB engine outperforms all classical engines where the TUR holds, which includes all Markovian classical engines. We note that there may exist non-Markovian classical engines which outperform the SSDB engine.

\begin{figure}
\centering
\includegraphics[width=\columnwidth]{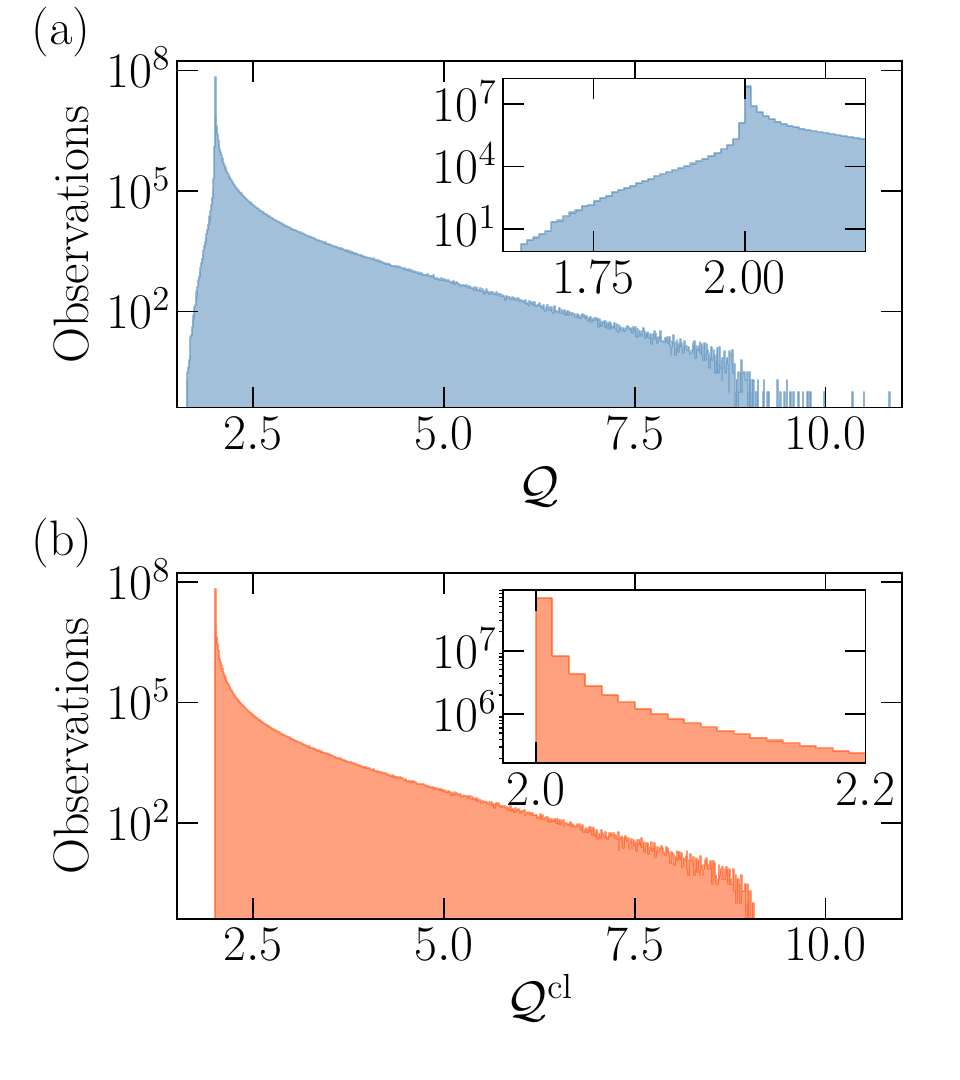}
\caption{Histograms of sampled values of (a) $\mathcal{Q}$ and (b) $\mathcal{Q}^\mathrm{cl}$ from Monte Carlo exploration of the parameter space.
Insets show the subset of the sampled data in the vicinity of the classical TUR limit.
The parameters are sampled from the uniform distributions $\gamma_\alpha\in[10^{-4};5]$, $n_\alpha\in[10^{-4};10]$, $\epsilon\in[10^{-4};1]$ and $\Delta\in[0;1]$, and the $10^8$ data points are arranged in bins of width 0.01.}
\label{FigMC}
\end{figure}

\begin{figure*}
	\centering
	\includegraphics[width=\textwidth]{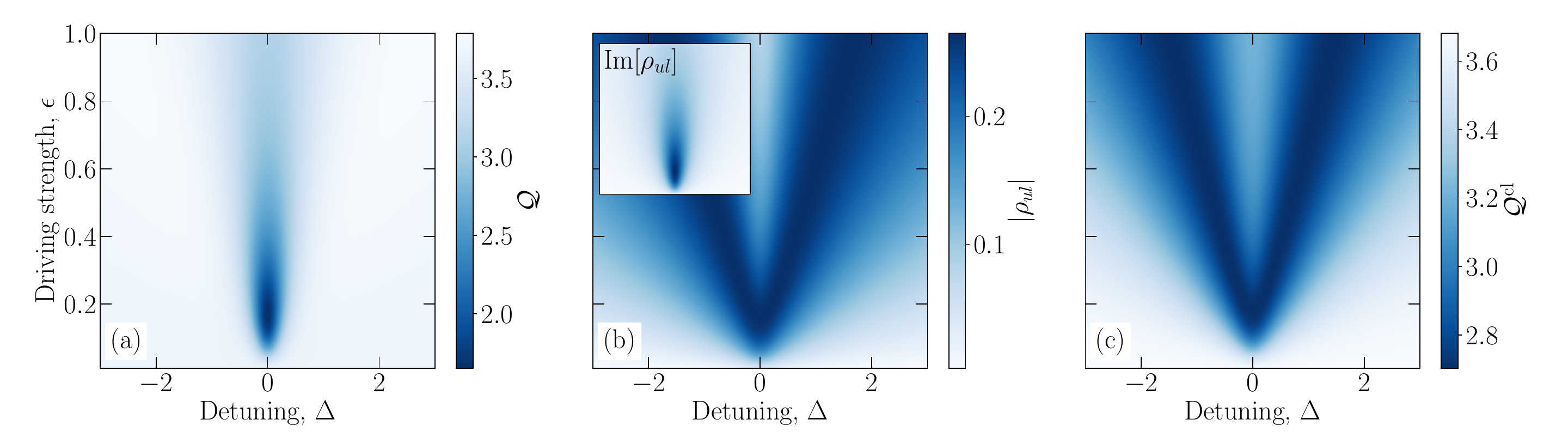}
	\caption{Heatmaps of (a) $\mathcal{Q}$ for the SSDB maser, (b) the off-diagonal matrix element $|\rho_{ul}|$ and its imaginary component (inset), and (c) $\mathcal{Q}^\mathrm{cl}$ for the classical reference system. The maser parameters are $\gamma_u = 2$, $n_u=0.027$, $\gamma_l=0.1$ and $n_l=5$.
		While $\mathcal{Q}$ shows a behavior similar to $\mathrm{Im}[\rho_{ul}]$, $\mathcal{Q}^\mathrm{cl}$ shows similar behavior to $|\rho_{ul}|$.}
	\label{FigHeat}
\end{figure*}

\textit{Monte Carlo exploration.---}
To explore the thermodynamic behavior of the model we systematically evaluated $\mathcal{Q}$ by Monte Carlo sampling over a region of the parameter space, see Fig.~\ref{FigMC}.  
Both the quantum and classical model operate close to the conventional TUR limit $\mathcal{Q}= 2$ in the great majority of sampled operation points.
Quantum violations of the conventional TUR are not uncommon, see the inset in Fig.~\ref{FigMC}~(a). On the other hand $\mathcal{Q}^\mathrm{cl}\ge 2$ is always satisfied as expected for systems based on classical rate equations. 
The range of $\mathcal{Q}$ stretches slightly wider to both smaller and larger values than $\mathcal{Q}^\mathrm{cl}$ due to the possibility of both quantum advantage and disadvantage, cf. Fig.~\ref{FigMC}~(a) and (b).

%\textit{The Thermodynamic Uncertainty Relation.---}
\textit{Bounds for $\mathcal{Q}^\mathrm{cl}$.---}
Based on Eqs.~(\ref{EqQturScovil}) and (\ref{EqFano}) the thermodynamic uncertainty can be cast into
\begin{equation}
\mathcal{Q} = \mathcal{Q}_\mathrm{pop} + \mathcal{Q}_\mathrm{tr}\,,
\end{equation}
where the first (population) term depends only on the bath populations and is the same for the quantum and the classical model
\begin{equation}
\mathcal{Q}_\mathrm{pop} =
\ln\left[\frac{n_l(n_u+1)}{n_u(n_l+1)}\right]
\frac{n_l(n_u+1)+n_u(n_l+1)}{n_l-n_u}\,.
\end{equation}
This term always adheres to the TUR, i.e., $\mathcal{Q}_\mathrm{pop}\geq 2$ (see supplemental material).
The transport term, which is essential for TUR violations, is given as
\begin{equation}
\mathcal{Q}_\mathrm{tr} =-\ln\left[\frac{n_l(n_u+1)}{n_u(n_l+1)}\right]
2\langle\dot{N}\rangle C\,.
\end{equation}
In the classical case, Eq.~\eqref{EqC} provides $C^\mathrm{cl}>0$ and the transport term is negative. As TUR is always satisfied here, we have the restriction $2\le \mathcal{Q}^\mathrm{cl}\le \mathcal{Q}_\mathrm{pop}$. Both inequalities fail in the quantum case, as we have the quantum advantage $C>C^\mathrm{cl}$ for $|\Delta|<\Gamma$, and $C$ can actually become negative for large values of $|\Delta|$. In agreement with Fig.~\ref{FigMC}, the thermodynamic uncertainty may thus take on a larger range of values in the quantum case.

\textit{Quantum coherence and TUR.---}
Previous work on TUR in quantum systems showed that quantum coherence can be responsible for TUR violations \cite{PtaszynskiPRB2018,AgarwallaPRB2018,RignonBretPRE2021}. 
In Fig.~\ref{FigHeat} we compare $\mathcal{Q}$ with the steady-state coherence $\rho_{ul} = \langle u|\rho|l\rangle$ between the lasing levels as a function of detuning and driving. We find particularly low values 
of $\mathcal{Q}$ for zero detuning  and moderate driving in Fig.~\ref{FigHeat}\,(a). This differs 
entirely from the coherence $|\rho_{ul}|$ shown in Fig.~\ref{FigHeat}\,(b), which exhibits a V-shaped 
ridge (see supplemental material). Surprisingly, a similar ridge is found in $\mathcal{Q}^\mathrm{cl}$, 
see Fig.~\ref{FigHeat}\,(c), which indicates that the absolute value $|\rho_{ul}|$ cannot be directly related 
to quantum effects in $\mathcal{Q}$. Furthermore, the V-shaped coherence pattern extends into regions 
$|\Delta|>\Gamma$, where quantum effects provide $\mathcal{Q}>\mathcal{Q}^\mathrm{cl}$, i.e., a quantum disadvantage.

On the other hand, the inset in Fig.~\ref{FigHeat}\,(b) shows, that 
$\mathrm{Im}[\rho_{ul}]$ might be a candidate to explain the TUR violations. In this context, we note that 
\begin{equation}
\langle\dot{N}\rangle=2\epsilon\mathrm{Im}[\rho_{ul}]
\label{EqMeanRate}
\end{equation}
as can be seen from the relation $\langle P\rangle = -\mathrm{Tr}[(\partial_t H)\rho]$. Nevertheless, a finite rate does not ensure the presence of coherence, as the classical model produces the exact same amount of power. 
Equation~\eqref{EqQadvantage} shows 
that any quantum advantage is proportional to $\langle\dot{N}\rangle$. Thus
it is rather the mean rate, which can be reproduced by the classical model, than the coherence itself, which is of relevance  here.

What distinguishes TUR in the quantum system from its classical counterpart is the variance, a quantity that goes beyond what is encoded in the steady state.
Hence, the steady-state coherence cannot fully capture what sets quantum-coherent dynamics apart from the classical system. A similar conclusion was reached in Ref.~\cite{medina:2021}.

\textit{Conclusion.---}
In this letter we studied the performance of the  Scovil \& Schulz-DuBois heat engine in terms of the thermodynamic uncertainty ${\cal Q}$.
From the analytical expressions we find that the quantum-coherent dynamics exhibits a quantum advantage compared to its classical counterpart if the detuning is less than the broadening, $|\Delta|<\Gamma$.
In cases of great quantum advantage, the maser violates the conventional TUR limit ${\cal Q}>2$.
Parameter points where the engine operates close to the conventional limit are prevalent and violations not uncommon. We also showed, that this engine
adheres to a recent formulation of a quantum TUR.

While quantum advantages and disadvantages over classical systems are related to the quantum-coherent dynamics, we cannot establish a direct relation between the behavior of the steady-state coherence and that of the thermodynamic uncertainty.
The thermodynamic uncertainty depends on the variance of the output power whose properties, unlike the mean, are not fully imprinted onto the state of the system.
This illustrates a need to introduce new quantifiers that measure the coherence present in the dynamics of a system going beyond the density matrix.

\textit{Acknowledgements.---}
We thank the Knut and Alice Wallenberg Foundation (project 2016.0089) and Nano\-Lund for financial support.
P.P.P. acknowledges funding from the European Union's Horizon 2020 research and innovation programme under the Marie Sk\l{}odowska-Curie Grant Agreement No. 796700, from the Swedish Research Council (Starting Grant 2020-03362), and from the Swiss National Science Foundation (Eccellenza Professorial Fellowship PCEFP2\_194268).

%\newpage
\bibliography{refs}

\clearpage
\widetext

\begin{center}
	\textbf{\large Supplemental Material: Violating the Thermodynamic Uncertainty Relation in the Three-Level Maser}
\end{center}

%%%%%%%%%% Merge with supplemental materials %%%%%%%%%%
%%%%%%%%%% Prefix a "S" to all equations, figures, tables and reset the counter %%%%%%%%%%
\renewcommand{\theequation}{S\arabic{equation}}
\renewcommand{\thefigure}{S\arabic{figure}}
%\renewcommand{\bibnumfmt}[1]{[S#1]}
%\renewcommand{\citenumfont}[1]{S#1}
%%%%%%%%%% Prefix a "S" to all equations, figures, tables and reset the counter %%%%%%%%%%

The following is a supplementary material detailing and deriving the equations presented in the main text.

\section{Steady State of the Three-Level Maser}
\label{AppSteady}
The Hamiltonian $H(t) = H_0 + V(t)$ consists of a bare term
\begin{equation}
H_0 = \omega_l\sigma_{ll}+\omega_u\sigma_{uu}+\omega_x\sigma_{xx}\,,
\end{equation}
and an external classical field
\begin{equation}
V(t) = \epsilon(\e^{\imai\omega_dt}\sigma_{lu}+e^{-\imai\omega_dt}\sigma_{ul})\,.
\end{equation}
To remove the time dependence of the Hamiltonian we transform to an appropriate rotating frame which simplifies the equation of motion.
We define rotated operators by the unitary transformation $A^\mathrm{rot} = U(t)AU^\dagger(t)$ with $U(t)=\e^{\imai X t}$ and $X = \omega_l\sigma_{ll}+(\omega_l+\omega_d)\sigma_{uu}+\omega_x\sigma_{xx}$. In this rotated frame, we obtain
\begin{equation}
\label{eq:mastereqrot}
\dot\rho = -\imai[\widetilde{H},\rho]+\sum_\alpha\left\{\gamma_\alpha(n_\alpha+1)\mathcal{D}_{\sigma_{\alpha x}}[\rho]+\gamma_\alpha n_\alpha\mathcal{D}_{\sigma_{x\alpha}}[\rho]\right\}\,,
\end{equation}
with the new Hamiltonian 
$\widetilde{H} = H^\mathrm{rot}-X = -\Delta\sigma_{uu}+\epsilon(\sigma_{ul}+\sigma_{lu})$ where we introduced the detuning parameter $\Delta = \omega_d - (\omega_u-\omega_l)$. Note, that the structure of the dissipators is not affected by this transformation.
From Eq.~\eqref{eq:mastereqrot}, we obtain the equations of motion for $\rho_{ij}=\langle i |\rho^\mathrm{rot}|j\rangle$
\begin{eqnarray}
\dot\rho_{xx}&=&\gamma_l n_l\rho_{ll}
+\gamma_u n_u\rho_{uu}-[(n_l+1)\gamma_l+(n_u+1)\gamma_u] \rho_{xx}\,,\\
\dot\rho_{uu}&=&\gamma_u (n_u+1)\rho_{xx}-\gamma_un_u\rho_{uu}
+\imai\epsilon (\rho_{ul}-\rho_{ul}^*)\label{Eqrhouu}\,,\\
\dot\rho_{ll}&=&\gamma_l (n_l+1)\rho_{xx}-\gamma_l n_l\rho_{ll}
-\imai\epsilon (\rho_{ul}-\rho_{ul}^*)\label{Eqrholl}\,,\\
\dot\rho_{ul}&=&\imai \Delta \rho_{ul}+\imai\epsilon(\rho_{uu}-\rho_{ll})-\Gamma\rho_{ul}\,, \label{Eqrhoul}
\end{eqnarray}
where $\Gamma=\frac12(\gamma_un_u+\gamma_ln_l)$ is the quantum decoherence rate.
In the steady state (superscript ss), Eq.~\eqref{Eqrhoul} provides
\begin{equation}
\rho^\textrm{ss}_{ul}=\frac{-\epsilon(\rho^\textrm{ss}_{uu}-\rho^\textrm{ss}_{ll})}{\Delta+\imai\Gamma}\,.\label{Eqrhoulstat}
\end{equation}
The steady-state populations are
\begin{equation}
\begin{split}
\rho^\textrm{ss}_{ll}&=\frac{\gamma_l\gamma_u n_u(n_l+1)+\gamma_c[\gamma_l(n_l+1)+\gamma_u( n_u+1)]}{ [\gamma_l(2n_l+1)+\gamma_c][\gamma_u(2n_u+1)+\gamma_c]-[\gamma_l(n_l+1)-\gamma_c][\gamma_u( n_u+1)-\gamma_c]}\,,\\
\rho^\textrm{ss}_{uu}&=\frac{\gamma_l\gamma_u n_l(n_u+1)+\gamma_c[\gamma_l(n_l+1)+\gamma_u( n_u+1)]}{ [\gamma_l(2n_l+1)+\gamma_c][\gamma_u(2n_u+1)+\gamma_c]-[\gamma_l(n_l+1)-\gamma_c][\gamma_u( n_u+1)-\gamma_c]}\,,\\
\rho^\mathrm{ss}_{xx}&=1-\rho_{uu}^\mathrm{ss}-\rho_{ll}^\mathrm{ss}\,,
\end{split}
\end{equation}
where the classical rate reads $\gamma_c = \frac{2\epsilon^2\Gamma}{\Delta^2+\Gamma^2}$.
The lasing populations provide
\[
\rho^\textrm{ss}_{uu}-\rho^\textrm{ss}_{ll}=\frac{\gamma_l\gamma_u (n_l-n_u)}
{2\gamma_c[3\Gamma+\gamma_u+\gamma_l]+\gamma_l\gamma_u(3n_l n_u+n_l+n_u)}\,,
\]
which is proportional to the occupation differences of the baths.
(Note that $n_l>n_u$ yields population inversion).
Inserting this into Eq.~(\ref{Eqrhoulstat}) we find
\begin{equation}
\rho^\textrm{ss}_{ul}=\frac{(-\Delta+\imai\Gamma)
	\epsilon  (n_l-n_u)}
{(\Delta^2+\Gamma^2)A+\epsilon^2B}
\quad\textrm{with }A=3n_l n_u+n_l+n_u
\quad\textrm{with }B= 2\Gamma[(3n_l+2)/\gamma_u+(3n_u+2)/\gamma_l]\, .
\end{equation}
By straightforward algebra, we find that $|\rho^\textrm{ss}_{ul}|$ has a ridge of maxima as a function of $\epsilon$ and $\Delta$ on the curve 
\begin{equation}
\epsilon^2=(\Delta^2+\Gamma^2)\frac{A}{B},
\end{equation}
with the constant peak value
\begin{equation}
|\rho^\textrm{ss}_{ul}|_\textrm{peak}=\frac{(n_l-n_u)}{2\sqrt{AB}}\,,
\end{equation}
as can be seen in Fig.~\ref{FigHeat}\,(b).

\section{Full Counting Statistics}
\label{AppFCS}
An analytical approach to determine the particle statistics in an open quantum system is provided by Full Counting Statistics (FCS), where counting fields are included in the master equation \cite{EspositoRevModPhys2009,SchallerBook2014}.
Let $\chi_u$ and $\chi_l$ be counting fields for the respective reservoirs.
The Lindblad master equation becomes
\begin{equation}
\dot{\rho}^\mathrm{rot} = -\imai[\widetilde{H},\rho^\mathrm{rot}] +\mathcal{L}_u^{\chi_u}[\rho^\mathrm{rot}]+\mathcal{L}_l^{\chi_l}[\rho^\mathrm{rot}]\,,
\end{equation}
with the modified Lindbladians
\begin{equation}
\mathcal{L}_\alpha^{\chi_\alpha}[\rho] =
\gamma_\alpha (n_\alpha+1)\mathcal{D}_{\sigma_{\alpha x}}^{\chi_\alpha}[\rho]+
\gamma_\alpha n_\alpha\mathcal{D}_{\sigma_{x\alpha}}^{-\chi_\alpha}[\rho]\,,
\end{equation}
and dissipators
\begin{equation}
\mathcal{D}_\sigma^\chi[\rho] = \e^{-\imai\chi}\sigma\rho\sigma^\dagger - \frac12\{\sigma^\dagger\sigma\rho + \rho\sigma^\dagger\sigma\}\,.
\end{equation}
If we reshape the density matrix into a state vector $\rho^\mathrm{rot}=(\rho_{xx},\rho_{uu},\rho_{ll},\mathrm{Re}[\rho_{ul}],\mathrm{Im}[\rho_{ul}])^T$ we summarize the Lindblad master equation as a matrix equation with the Liouvillian supermatrix $\mathcal{L}(\chi_u,\chi_l)$ \cite{BreuerBook2002}
\begin{equation}
\dot{\rho}^\mathrm{rot} = \mathcal{L}(\chi_u,\chi_l)\rho^\mathrm{rot}\,.
\end{equation}
The full Liouvillian supermatrix with counting fields is
\begin{equation}
\mathcal{L}(\chi_u,\chi_l) = \left[
\begin{matrix}
-\gamma_u(n_u+1)-\gamma_l(n_l+1) & \gamma_un_u\e^{\imai\chi_u} & \gamma_ln_l\e^{\imai\chi_l} & 0 & 0\\
\gamma_u(n_u+1)\e^{-\imai \chi_u} & -\gamma_un_u & 0 & 0 & -2\epsilon\\
\gamma_l(n_l+1)\e^{-\imai \chi_l} & 0 & -\gamma_ln_l & 0 & 2\epsilon\\
0 & 0 & 0 & -\Gamma & \Delta\\
0 & \epsilon & -\epsilon & -\Delta & -\Gamma\\
\end{matrix}
\right]\,.
\label{EqFCSLiouvillian}
\end{equation}
In the limit $\chi_u,\chi_l\rightarrow0$ this reduces to the original Liouvillian supermatrix for proper time evolution. As shown in Sec.~\ref{app:counting}, it is sufficient for our purposes to count the quanta exchanged with bath $u$. Therefore, we set $\chi_l=0$ and $\chi_u=\chi$ in the following.

In the large time limit the $k$'th cumulant of the integrated number of quanta emitted into reservoir $\alpha$
over a time window $t$ can be determined by \cite{SchallerBook2014}
\begin{equation}
\label{eq:cumulants}
C^k(t) =\left.(\imai\partial_{\chi})^k\left[\zeta(\chi)t + c(\chi)\right]\right|_{\chi=0}\,,
\end{equation}
where $\zeta(\chi)$ is the eigenvalue of $\mathcal{L}(\chi)\equiv\mathcal{L}(\chi,0)$ with the largest real part and $c(\chi)$ is a polynomial depending on the eigenvectors of $\mathcal{L}(\chi)$.
The first and second cumulants correspond to the mean and variance of the integrated particle current respectively
\begin{equation}
\langle\dot{N}\rangle \simeq \left. \imai \partial_{\chi}\zeta(\chi)\right|_{\chi = 0}\,,\hspace{1.5cm}
\mathrm{var}(\dot{N}) \simeq\left. -\partial_{\chi}^2\zeta(\chi)\right|_{\chi = 0}\,,
\label{EqTURCumulant}
\end{equation}
where we have dropped the term in Eq.~\eqref{eq:cumulants} that does not grow in time.

To determine the mean and variance from the derivatives analytically we follow the method outlined in Ref.~\cite{BrudererNJP2014}.
Consider the characteristic polynomial of $\mathcal{L}(\chi)$
\begin{equation}
\sum_n a_n \zeta^n = 0\,,
\end{equation}
where the terms $a_n$ are functions of $\chi$.
Define
\begin{equation}
a_n' = \imai\partial_{\chi} a_n|_{\chi=0},
\quad a_n'' = (\imai\partial_{\chi})^2a_n|_{\chi=0} = -\partial_{\chi_u}^2a_n|_{\chi=0}\,,
\end{equation}
and similarly for $\zeta$.
We determine the derivative of the polynomial equation in $\chi$
\begin{equation}
\left[\imai\partial_{\chi}\sum_n a_n\zeta^n\right]_{\chi=0} = \sum_n[a_n'+(n+1)a_{n+1}\zeta']\zeta^n(0) = 0\,.
\label{EqFCSfirst}
\end{equation}
Continuing with the second derivative we find
\begin{equation}
\left[(\imai\partial_\chi)^2\sum_na_n\zeta^n\right]_{\chi=0} =
\sum_n[a_n''+2(n+1)a_{n+1}'\zeta'+(n+1)a_{n+1}\zeta''+(n+1)(n+2)a_{n+2}\zeta'^2]\zeta^n(0)=0\,.
\label{EqFCSsecond}
\end{equation}
We assume that the system has a unique steady state for which $\zeta(0)=0$.
We know that the zeroth term $\zeta^0=1$ must vanish, hence Eq.~\eqref{EqFCSfirst} indicates
\begin{equation}
a_0'+a_1\zeta'=0\,,
\end{equation}
which provides the current
\begin{equation}
\langle\dot{N}\rangle = \zeta' = -\frac{a_0'}{a_1}\,.
\label{EqFCSmean}
\end{equation}
We obtain the variance similarly from Eq.~\eqref{EqFCSsecond}
\begin{equation}
\mathrm{var}(\dot{N}) = \zeta'' =-\frac{a_0''+2\langle\dot{N}\rangle(a_1'+a_2\langle\dot{N}\rangle)}{a_1} =  2\frac{a_0'a_1a_1'-a_0'^2a_1'}{a_1^3}-\frac{a_0''}{a_1}.
\label{EqFCSvariance}
\end{equation}
The expressions \eqref{EqFCSmean} and \eqref{EqFCSvariance} hold for all systems with Lindblad dynamics assuming a unique steady state.

The Liouvillian given in Eq.~\eqref{EqFCSLiouvillian} results in the parameters
\begin{equation}
\begin{split}
a_0'=&-2\epsilon^{2} \gamma_{l} \gamma_{u}(n_l-n_u)\Gamma\,,\\
a_0'' =& -2\epsilon^{2}\gamma_u\gamma_l\Gamma \left(2 n_{l} n_{u} + n_{l} + n_{u}\right)\,,\\
a_1 =& \gamma_l\gamma_u[\Delta^2+\Gamma^2](3n_ln_u+n_u+n_l)
+4\epsilon^{2}\Gamma\left[3\Gamma+\gamma_l+\gamma_u \right]\,,\\
a_1' =&- 2\epsilon^2\gamma_u\gamma_l(n_l-n_u)\,,\\
a_2 =& [\Delta^{2}+\Gamma^2+4\epsilon^2] \left(4\Gamma + \gamma_{l} + \gamma_{u}\right) + 2 \Gamma \gamma_{l} \gamma_{u} \left(3 n_{l} n_{u} + n_{l} + n_{u}\right)\,,
\end{split}
\end{equation}
where $\Gamma=\frac12(\gamma_un_u+\gamma_ln_l)$ is the decoherence rate.
This provides the current
\begin{equation}
\langle\dot{N}\rangle = \frac
{2\gamma_u\gamma_l\epsilon^{2}(n_l-n_u)\Gamma}
{[\Delta^2+\Gamma^2]\gamma_u\gamma_l(3n_ln_u+n_u+n_l)+4\epsilon^{2}
	\Gamma\left[3\Gamma+\gamma_u+\gamma_l \right]},
\label{EqFCScurrent}
\end{equation}
and the Fano factor
\begin{equation}
F(\dot{N}) = \frac{\mathrm{var}(\dot{N})}{\langle\dot{N}\rangle}=\frac{a_0''}{a_0'}+2\langle\dot{N}\rangle\left[\frac{a_1'}{a_0'}-\frac{a_2}{a_1}\right]=\frac{n_u(n_l+1)+n_l(n_u+1)}{n_l-n_u}+2\langle\dot{N}\rangle
\left[\Gamma^{-1}-\frac{a_2}{a_1}\right].
\end{equation}
We introduce the population and transport terms of the Fano factor, respectively,
\begin{equation}
F_\mathrm{pop} = \frac{n_u(n_l+1)+n_l(n_u+1)}{n_l-n_u},\quad
F_\mathrm{tr} = -2\langle\dot{N}\rangle\left[\frac{a_2}{a_1}-\Gamma^{-1}\right],
\label{EqFCSfano}
\end{equation}
for which $F = F_\mathrm{pop}+F_\mathrm{tr}$.
Finally, from the Fano factors we define the thermodynamic uncertainty similarly.
Per Eq.~\eqref{EqQturScovil}
\begin{equation}
\mathcal{Q} = \ln\left[\frac{n_l(n_u+1)}{n_u(n_l+1)}\right]F_\mathrm{pop}
+\ln\left[\frac{n_l(n_u+1)}{n_u(n_l+1)}\right]F_\mathrm{tr}=\mathcal{Q}_\mathrm{pop}+\mathcal{Q}_\mathrm{tr},
\label{EqFCSqtur}
\end{equation}
where $\mathcal{Q}_\mathrm{pop}$ is the population term and $\mathcal{Q}_\mathrm{tr}$ the transport term.
Using the inequalities $a/(a-b)\ln(a/b)\geq1$ and $b/(a-b)\ln(a/b)\geq1$ we can show that
\begin{equation}
\mathcal{Q}_\mathrm{pop}=\ln\left[\frac{n_l(n_u+1)}{n_u(n_l+1)}\right]\frac{n_u(n_l+1)+n_l(n_u+1)}{n_l-n_u}\geq2\,.
\label{EqFCSqturpop}
\end{equation}
The population term adheres to the TUR limit for all parameters, hence, the transport term is essential for TUR violations.
For some parameters $\mathcal{Q}_\mathrm{tr}$ is positive and decreases the engine precision.

\section{Connecting FCS to the average and variance of power}
	\label{app:counting}
	In this section, we show that in the long time limit, counting the heat quanta exchanged with the bath $u$ is sufficient for determining the statistics of all heat and work flows up to the variance. We start by showing that the statistics of the heat quanta exchanged with the two baths are identical. To this end, we consider the unitary superoperator
	\begin{equation}
	\mathcal{U} {\rho} = e^{-\frac{\imai}{2}\chi_l(\sigma_{uu}+\sigma_{ll})}{\rho}e^{-\frac{\imai}{2}\chi_l(\sigma_{uu}+\sigma_{ll})},\hspace{1cm}\mathcal{U}^\dagger {\rho} = e^{\frac{\imai}{2}\chi_l(\sigma_{uu}+\sigma_{ll})}{\rho}e^{\frac{\imai}{2}\chi_l(\sigma_{uu}+\sigma_{ll})}.
	\end{equation}
	A straightforward calculation shows that
	\begin{equation}
	\label{eq:liouvunit}
	\mathcal{U}^\dagger\mathcal{L}(\chi_u,\chi_l)\mathcal{U} = \mathcal{L}(\chi_u+\chi_l,0).
	\end{equation}
	In the long-time limit, the statistics are fully determined by the eigenvalues of the Liouvillian, which do not change under a unitary transformation. Equation \eqref{eq:liouvunit} thus implies that the statistics of the heat quanta exchanged with the two baths are identical and that it is sufficient to consider a single counting field in the long-time limit.

To connect the mean and variance of the exchanged heat quanta to the power operator, we consider the unitary superoperator
	\begin{equation}
	\mathcal{V} {\rho} = e^{-\frac{\imai}{2}\chi\sigma_{uu}}{\rho}e^{-\frac{\imai}{2}\chi\sigma_{uu}}.
	\end{equation}
	Transforming the Liouvillian then results in
	\begin{equation}
	\label{eq:Liouvtr}
	\hat{\mathcal{L}}(\chi)\rho=\mathcal{V}^\dagger\mathcal{L}(\chi)\mathcal{V}{\rho} = -\imai \widetilde{H}(\chi)\hat{\rho} +\imai {\rho}\widetilde{H}(-\chi)+\mathcal{L}_u[\rho]+\mathcal{L}_l[\rho],
	\end{equation}
	with
	\begin{equation}
	\label{eq:hamchi}
	\widetilde{H}(\chi) = -\Delta \sigma_{uu}+\epsilon\left(\sigma_{ul}e^{\frac{\imai}{2}\chi}+\sigma_{lu}e^{\frac{-\imai}{2}\chi}\right).
	\end{equation}
	We may now compute $\langle \dot{N}\rangle$ and $\mathrm{var}(\dot{N})$ from $\hat{\mathcal{L}}$. Following the methods outlined in Ref.~\cite{flindt:2010}, we find for the average
	\begin{equation}
	\label{eq:avgpow}
	\langle \dot{N}\rangle={\rm Tr}\left[\hat{\mathcal{L}}'\rho^{\rm ss}\right] = \frac{1}{\omega_d}{\rm Tr}\left[P^{\rm rot}\rho^{\rm ss}\right],
	\end{equation}
	where $\hat{\mathcal{L}}'=i\partial_{\chi}\hat{{\mathcal{L}}}(\chi)|_{\chi=0}$ and the power operator in the rotating frame reads
	\begin{equation}
	\label{eq:powerrot}
	P^{\rm rot}=U(t)[-\partial_t H]U^\dagger(t).
	\end{equation}
	This confirms that $\langle \dot{N}\rangle=\langle P\rangle/\omega_d$ denotes the number of photons emitted into the driving field.

The variance can be written as \cite{flindt:2010} (dropping a term that vanishes for our Liouvillian)
	\begin{equation}
	\label{eq:var}
	\mathrm{var}(\dot{N}) = 2{\rm Tr}\left[\hat{\mathcal{L}}'\mathcal{L}^{\rm D}\hat{\mathcal{L}}' \rho^{\rm ss}\right],
	\end{equation}
	with the Drazin inverse of $\mathcal{L}\equiv\mathcal{L}(0)$ given by \cite{mandal:2016,scandi:2019}
	\begin{equation}
	\label{eq:drazininv}
	\mathcal{L}^{\rm D} = -\int_{0}^{\infty}d\tau e^{\mathcal{L}\tau}(\mathcal{I}-\mathcal{P}),
	\end{equation}
	where $\mathcal{I}$ denotes the identity and $\mathcal{P}$ projects onto the null space of the Liouvillian, i.e., $\mathcal{P}\rho=\rho^{\rm ss}{\rm Tr}[\rho]$, see also Eq.~\eqref{eq:pproj} below.
	Equation \eqref{eq:var} can the expressed as
	\begin{equation}
	\label{eq:var2}
	\mathrm{var}(\dot{N}) =\frac{1}{\omega_d^2}\int_{0}^\infty d\tau{\rm Tr}\left[\delta P^{\rm rot}e^{\mathcal{L}\tau}\{\delta P^{\rm rot},\rho^{\rm ss}\}\right] = \frac{1}{\omega_d^2}\int_{-\infty}^{\infty}\frac{1}{2}\langle \{\delta P(t+\tau),\delta P(t)\}\rangle,
	\end{equation}
	where $\delta P^{\rm rot}=P^{\rm rot}-{\rm Tr}[P^{\rm rot}\rho^{\rm ss}]$. To obtain the last equality in Eq.~\eqref{eq:var2}, which connects $\mathrm{var}(\dot{N})$ to the standard low frequency power fluctuations, we employed standard equalities for two-point correlation functions in open quantum systems \cite{gardiner:book}.

\section{A Classical Formulation of the Three-Level Maser}
\label{AppClassic}

To understand the quantum nature of the SSDB maser we construct a classical equivalent system for reference.
We omit the Hamiltonian from the master equation and add the Lindblad jump operators $\sigma_{ul}$ and $\sigma_{lu}$ with the coupling rate $\gamma_c$.
Any coherence vanishes in this formulation and we consider the vectorized density matrix $\rho^\mathrm{cl}=(\rho_{xx},\rho_{uu},\rho_{ll})^T$ with the Liouvillian
\begin{equation}
\mathcal{L}^\mathrm{cl}=
\left[\begin{matrix}- \gamma_{l} \left(n_{l} + 1\right) - \gamma_{u} \left(n_{u} + 1\right) & \gamma_{u} n_{u} e^{i \chi} & \gamma_{l} n_{l}\\\gamma_{u} \left(n_{u} + 1\right) e^{- i \chi} & - \gamma_{c} - \gamma_{u} n_{u} & \gamma_{c}\\\gamma_{l} \left(n_{l} + 1\right) & \gamma_{c} & - \gamma_{c} - \gamma_{l} n_{l}\end{matrix}\right]\,.
\end{equation}
Similarly to Sec~\ref{AppFCS} we determine the polynomial factors with respective derivatives
\begin{equation}
\begin{split}
a_0'=&- \gamma_{c} \gamma_{l} \gamma_{u} \left(n_{l} - n_{u}\right)\,,\\
a_0'' = & - \gamma_{c} \gamma_{l} \gamma_{u} \left(2n_{l} n_{u} + n_{l} + n_{u}\right)\,,\\
a_1 =& 2\gamma_c[3\Gamma+\gamma_l+\gamma_u]+\gamma_l\gamma_u(3n_ln_u+n_l+n_u)\,,\\
a_1' =& 0\,,\\
a_2 =& 2\gamma_c+4\Gamma+\gamma_l+\gamma_u\,,
\end{split}
\end{equation}
where $\Gamma = \frac12(\gamma_un_u+\gamma_ln_l)$ is the quantum decoherence rate.
The classical steady-state current reads
\begin{equation}
\langle\dot{N}^\mathrm{cl}\rangle = -\frac{a_0'}{a_1} =
\frac{\gamma_{c}\gamma_u\gamma_l \left(n_{l} - n_{u}\right)}
{2\gamma_c[3\Gamma+\gamma_u+\gamma_l]+\gamma_l\gamma_u(3n_ln_u+n_l+n_u)}\,.
\end{equation}
Comparing with the quantum current \eqref{EqFCScurrent} we note that the two currents coincide $\langle\dot{N}^\mathrm{cl}\rangle = \langle\dot{N}\rangle$ when the coupling rate is defined as
\begin{equation}
\gamma_c = \frac{2\epsilon^2\Gamma}{\Delta^2+\Gamma^2}\,.
\end{equation}

Next, we determine the classical Fano factor
\begin{equation}
\begin{split}
F(\dot{N}^\mathrm{cl}) &=\frac{a_0''}{a_0'}-2\langle\dot{N}^\mathrm{cl}\rangle\left[\frac{a_2}{a_1}-\frac{a_1'}{a_0'}\right]\\
&= \frac{n_l(n_u+1)+n_u(n_l+1)}{n_l-n_u}-2\langle\dot{N}^\mathrm{cl}\rangle
\frac{2\gamma_c+4\Gamma+\gamma_l+\gamma_u}
{2\gamma_c[3\Gamma+\gamma_l+\gamma_u]+\gamma_l\gamma_u(3n_ln_u+n_l+n_u)}\,.\\
\end{split}
\end{equation}
Similarly to Sec~\ref{AppFCS}, we can write the Fano factor as the sum $F^\mathrm{cl} = F_\mathrm{pop} + F_\mathrm{tr}^\mathrm{cl}$ where the population term is identical to its quantum counterpart \eqref{EqFCSfano} and
\begin{equation}
F_\mathrm{tr}^\mathrm{cl}=
-2\langle\dot{N}^\mathrm{cl}\rangle
\frac{2\gamma_c+4\Gamma+\gamma_l+\gamma_u}
{2\gamma_c[3\Gamma+\gamma_l+\gamma_u]+\gamma_l\gamma_u(3n_ln_u+n_l+n_u)}\,,
\end{equation}
defines the classical Fano transport term.
Finally, we define the classical thermodynamic uncertainty
\begin{equation}
\mathcal{Q}^\mathrm{cl} =
\ln\left[\frac{n_l(n_u+1)}{n_u(n_l+1)}\right]F_\mathrm{pop}+
\ln\left[\frac{n_l(n_u+1)}{n_u(n_l+1)}\right]F_\mathrm{tr}^\mathrm{cl}= \mathcal{Q}_\mathrm{pop} + \mathcal{Q}^\mathrm{cl}_\mathrm{tr}\,.
\end{equation}
Similarly to the Fano factor, $\mathcal{Q}^\mathrm{cl}$ contains a population term identical to its quantum counterpart \eqref{EqFCSqturpop}.
As described in Sec.~\ref{AppFCS} this is restricted to values above $\mathcal{Q}_\mathrm{pop}\geq2$.
Considering the classical transport term $\mathcal{Q}_\mathrm{tr}^\mathrm{cl}$ we note that it can never be positive, unlike its quantum counterpart $\mathcal{Q}_\mathrm{tr}$.
Hence, $\mathcal{Q}_\mathrm{pop}$ is an upper bound for the classical thermodynamic uncertainty.

\section{Quantum Thermodynamic Uncertainty Relation}
\label{AppHasegawa}

A quantum formulation of TUR was recently suggested which for Lindblad master equations in our notation reads \cite{HasegawaPRL2020}
\begin{equation}
\frac{\mathrm{var}(\dot{N})}{\langle \dot{N}\rangle^2}\geq \frac{1}{\Upsilon+\Psi}\,,
\label{EqHasegawa2020}
\end{equation}
where the quantum dynamical activity is the average rate of transitions in the steady-state and reads
\begin{equation}
\Upsilon  = \sum_{\alpha\in\{u,l\}}\gamma_\alpha(1+n_\alpha)\rho^\mathrm{ss}_{xx}+\gamma_\alpha n_\alpha \rho^\mathrm{ss}_{\alpha},
\end{equation}
and the coherent-dynamics contribution reads
\begin{equation}
\Psi = -4\mathrm{Tr}[\mathcal{K}_1\mathcal{L}^{\rm D}\mathcal{K}_2[\rho^\mathrm{ss}]+\mathcal{K}_2\mathcal{L}^{\rm D}\mathcal{K}_1[\rho^\mathrm{ss}]]\,,
\end{equation}
where $\mathcal{K}_1[\rho] = -\imai\widetilde{H}\rho+\frac12\sum_\alpha\mathcal{L}_\alpha[\rho]$
and
$\mathcal{K}_2[\rho] = \imai\rho\widetilde{H}+\frac12\sum_\alpha\mathcal{L}_\alpha[\rho]$,
and $\mathcal{L}^{\rm D}$ is the Drazin inverse of the Liouvillian $\mathcal{L}=\mathcal{K}_1+\mathcal{K}_2$. 
In Sec~\ref{AppFCS} we exploited the fact that the Lindblad master equation preserves hermiticity of the density matrix in order to write the master equation in a reduced basis of the populations and upper-diagonal coherence terms.
However, the superoperators $\mathcal{K}$ do not individually preserve hermiticity of the density matrix, hence, the reduced basis employed in Sec.~\ref{AppFCS} becomes inconvenient.
For the vectorized density matrix in the basis $\rho^\mathrm{rot} = (\rho_{xx},\rho_{uu},\rho_{ll},\rho_{ul},\rho_{lu})^T$ the $\mathcal{K}$-matrices read
\begin{equation}
\mathcal{K}_1 =
\left[\begin{matrix}- \frac{\gamma_{l} \left(n_{l} + 1\right)}{2} - \frac{\gamma_{u} \left(n_{u} + 1\right)}{2} & \frac{\gamma_{u} n_{u}}{2} & \frac{\gamma_{l} n_{l}}{2} & 0 & 0\\\frac{\gamma_{u} \left(n_{u} + 1\right)}{2} & i \Delta - \frac{\gamma_{u} n_{u}}{2} & 0 & 0 & - i \epsilon\\\frac{\gamma_{l} \left(n_{l} + 1\right)}{2} & 0 & - \frac{\gamma_{l} n_{l}}{2} & - i \epsilon & 0\\0 & 0 & - i \epsilon & i \Delta - \frac{\gamma_{u} n_{u}}{2} & 0\\0 & - i \epsilon & 0 & 0 & - \frac{\gamma_{l} n_{l}}{2}\end{matrix}\right]\,,
\end{equation}
and
\begin{equation}
\mathcal{K}_2 =
\left[\begin{matrix}- \frac{\gamma_{l} \left(n_{l} + 1\right)}{2} - \frac{\gamma_{u} \left(n_{u} + 1\right)}{2} & \frac{\gamma_{u} n_{u}}{2} & \frac{\gamma_{l} n_{l}}{2} & 0 & 0\\\frac{\gamma_{u} \left(n_{u} + 1\right)}{2} & - i \Delta - \frac{\gamma_{u} n_{u}}{2} & 0 & i \epsilon & 0\\\frac{\gamma_{l} \left(n_{l} + 1\right)}{2} & 0 & - \frac{\gamma_{l} n_{l}}{2} & 0 & i \epsilon\\0 & i \epsilon & 0 & - \frac{\gamma_{l} n_{l}}{2} & 0\\0 & 0 & i \epsilon & 0 & - i \Delta - \frac{\gamma_{u} n_{u}}{2}\end{matrix}\right]\,.
\end{equation}
The Drazin inverse can be computed with the help of the Moore-Penrose pseudoinverse $\mathcal{L}^{\rm MP}=(\mathcal{L}^\dagger\mathcal{L})^{-1}\mathcal{L}^\dagger$ as
\begin{equation}
\mathcal{L}^{\rm D} = (\mathcal{I}-\mathcal{P})\mathcal{L}^{\rm MP}(\mathcal{I}-\mathcal{P})\,,
\end{equation}
where $\mathcal{I}_{ij}=\delta_{ij}$, with $\delta_{ij}$ denoting the Kronecker delta, and $\mathcal{P}$ the projection operator
\begin{equation}
\label{eq:pproj}
\mathcal{P} = [\rho^\mathrm{ss}, \rho^\mathrm{ss}, \rho^\mathrm{ss}, \vec{0}, \vec{0}]\,,
\end{equation}
where the columns are repeated instances of the vectorized density matrix and a vectorized zero state $\vec{0}=(0,0,0,0,0)^\dagger$.

Rearranging \eqref{EqHasegawa2020} we obtain the time-differentiated quantum TUR 
\begin{equation}
\mathcal{Q}=
\sigma\frac{\mathrm{var}(\dot{N})}{\langle\dot{N}\rangle^2}\geq \frac{\sigma}{\Upsilon+\Psi} \equiv \mathcal{B}\,,
\end{equation}
as used in Fig.~\ref{FigProbe}.

\end{document}